\documentclass[11pt]{article}
\usepackage{amsmath,amssymb,color}

\def\ben{\begin{equation}}
\def\een{\end{equation}}

\def\nn{\nonumber} \def\bd{\begin{document}} \def\ed{\end{document}}
\def\ds{\documentstyle} \let\fr=\frac \let\bl=\bigl \let\br=\bigr
\let\Br=\Bigr \let\Bl=\Bigl
\let\bm=\bibitem
\let\na=\nabla
\let\pa=\partial \let\ov=\overline
\newcommand{\be}{\begin{equation}}
\newcommand{\ee}{\end{equation}}
\def\ba{\begin{array}}
\def\ea{\end{array}}
\def\ft#1#2{{\textstyle{\frac{\scriptstyle #1}{\scriptstyle #2} } }}
\def\fft#1#2{{\frac{#1}{#2}}}
\def\del{\partial}
\def\vp{\varphi}
\def\sst#1{{\scriptscriptstyle #1}}
\def\oneone{\rlap 1\mkern4mu{\rm l}}
\def\td{\tilde}
\def\wtd{\widetilde}
\def\ie{{\it i.e.\ }}
\def\dalemb#1#2{{\vbox{\hrule height .#2pt
        \hbox{\vrule width.#2pt height#1pt \kern#1pt
                \vrule width.#2pt}
        \hrule height.#2pt}}}
\def\square{\mathord{\dalemb{6.8}{7}\hbox{\hskip1pt}}}
\def\i{{\rm i}}
\newcommand{\ho}[1]{$\, ^{#1}$}
\newcommand{\hoch}[1]{$\, ^{#1}$}
\newcommand{\bea}{\setlength\arraycolsep{2pt} \begin{eqnarray}}
\newcommand{\eea}{\end{eqnarray}}
\newcommand{\ra}{\rightarrow}
\newcommand{\lra}{\longrightarrow}
\newcommand{\Lra}{\Leftrightarrow}
\newcommand{\bp}{\tilde \beta^\prime}
\newcommand{\tr}{{\rm tr} }
\newcommand{\Tr}{{\rm Tr} }
\def\0{{\sst{(0)}}}
\def\1{{\sst{(1)}}}
\def\2{{\sst{(2)}}}
\def\3{{\sst{(3)}}}
\def\4{{\sst{(4)}}}
\def\5{{\sst{(5)}}}
\def\6{{\sst{(6)}}}
\def\7{{\sst{(7)}}}
\def\8{{\sst{(8)}}}
\def\m{{\sst{(m)}}}
\def\n{{\sst{(n)}}}
\def\cA{{{\cal A}}}
\def\cB{{{\cal B}}}
\def\cF{{{\cal F}}}
\def\cG{{{\cal G}}}
\def\cH{{{\cal H}}}
\def\tV{\widetilde V}
\def\tW{\widetilde W}
\def\tH{\widetilde H}
\def\tE{\widetilde E}
\def\tF{\widetilde F}
\def\tA{\widetilde A}
\def\im{{{\rm i}}}
\def\tY{{{\wtd Y}}}
\def\ep{{\epsilon}}
\def\vep{{\varepsilon}}
\def\bD{{{\bar D}}}
\def\R{{{\mathbb R}}}
\def\C{{{\mathbb C}}}
\def\H{{{\mathbb H}}}
\def\CP{{{\mathbb C}{\mathbb P}}}
\def\RP{{{\mathbb R}{\mathbb P}}}
\def\Z{{{\mathbb Z}}}
\def\bA{{{\mathbb A}}}
\def\bB{{{\mathbb B}}}
\def\bC{{{\mathbb C}}}
\def\bD{{{\mathbb D}}}
\def\bE{{{\mathbb E}}}
\def\bZ{{{\mathbb Z}}}
\def\Re{{{\frak{Re}}}}
\def\Im{{{\frak{Im}}}}
\def\cosec{{\,\hbox{cosec}\,}}
\def\Gm{{\Gamma_{\!\! -}}}
\def\Gp{{\Gamma_{\!\! +}}}
\def\stan{{standard }}
\def\nonstan{{supernumerary }}
\def\p{{\partial}}
\def\kdel#1{{\fft{\del}{\del#1}}}

\def\bog{{Bogomolny }}
\def\om{{\omega}}

\newcommand{\nnr}{\nonumber \\}
\newcommand{\pd}{\partial}
\newcommand{\ud}{\textrm{d}}
\newcommand{\dTH}{T^{\prime \, 0}_\textrm{H}}
\newcommand{\dOi}{\Omega^{\prime \, 0}_i}
\newcommand{\bx}{{\bf x}}

 \newcommand{\bpsi}{\bar\psi}
 \newcommand{\cL}{{\cal L}}
 \newcommand{\cR}{{\cal R}}
 \newcommand{\cT}{{\cal T}}
 \def\bra#1{\left\langle#1\right|}
 \def\ket#1{\left|#1\right\rangle}
 \def\brk#1{\left\langle#1\right\rangle}

\begin{document}

\begin{flushright}
\hfill{USTC-ICTS-10-04}

\end{flushright}

\vspace{25pt}
\begin{center}
{\large {\bf Calabi-Yau $(p+1)$-folds from $p$-folds}}

\vspace{15pt}

H. L\"u\hoch{\dagger\ddagger} and Zhao-Long Wang\hoch{\star}

\vspace{10pt}

\hoch{\dagger}{\it China Economics and Management Academy\\
Central University of Finance and Economics, Beijing, 100081}

\vspace{10pt}

\hoch{\ddagger}{\it Institute for Advanced Study, Shenzhen
University, Nanhai Ave 3688, Shenzhen 518060}

\vspace{10pt}

\hoch{\star}{\it Interdisciplinary Center for Theoretical Study,\\
University of Science and Technology of China, Hefei, Anhui 230026}

\vspace{40pt}

\underline{ABSTRACT}
\end{center}

We establish the general formalism for constructing metrics of
Calabi-Yau $(p+1)$-folds in terms of that of a $p$-fold by adding a
complex-line bundle.  We present a few explicit low-lying examples.
We further consider holomorphic linearization and obtain the
six-dimensional analogue of the Gibbons-Hawking instanton.  Whilst
the K\"ahler potential for the Gibbons-Hawking instanton is given by
the harmonic function of a three-dimensional flat space, for the
generalized solution it is related to the harmonic functions of
certain three-dimensional non-flat spaces that are direct products
of $\R$ and two-dimensional K\"ahler spaces.

\vspace{15pt}

\thispagestyle{empty}





\newpage

\section{Introduction}

      Six-dimensional Calabi-Yau (CY) manifolds
\cite{calabi1,calabi2,styau} play an important role in string
theory, since they provide natural internal compactifying spaces,
giving rise to four-dimensional theories that preserve one quarter
of the ten-dimensional supersymmetry. It is thus of great interest
to construct explicit metrics on Calabi-Yau manifolds.  It can be
argued however that such metrics are typically either non-compact or
singular. Recently metrics on Calabi-Yau 3-folds were constructed
from a generic hyper-K\"ahler space in $D=4$ by adding a
complex-line bundle \cite{lpw}. (See also \cite{sant}.) In this
follow-up paper we generalize the result to give the general
recursive relation of constructing the metrics of Calabi-Yau
$(p+1)$-folds in terms of a $p$-fold.  The metrics of Calabi-Yau
$(p+1)$-folds are obtained by deforming the K\"ahler potential of the Calabi-Yau
$p$-fold with a function $G$ that satisfies a single basic equation. We establish the general
formalism in section 2. In section 3, we present a few explicit
low-lying examples including $p=1,2$ and 3.

   A major addition in this paper is the systematical study of the
holomorphic linearization suggested in \cite{fay,sant}.  The basic
equations for the function $G$ in the construction are
increasingly non-linear and more difficult to solve for larger $p$.
The case for $p=1$ is special, leading to a K\"ahler potential that
depends on a harmonic function of three-dimensional flat space.  The
resulting metric is the Gibbons-Hawking instanton.  In section 4, we
consider holomorphic linearization for $p=2$, where the non-linear
terms in the basic equation of $G$
vanish.  This restricts the possible CY2 bases, which turn out to be
determined by the solutions of the two-dimensional Liouville
equation. For certain choices of special solutions, the function $G$
becomes a harmonic function of a certain three-dimensional non-flat
space that is a direct product of $\R$ and a two-dimensional
K\"ahler space. We conclude our paper in section 5. In the appendix,
we present a rather general solution that unifies the two special
examples presented in section 4.

\section{The construction}

   In this section, we present the metric construction for the CY
$(p+1)$-folds from CY $p$-folds.  This is the generalization of
\cite{lpw,sant} where $p=2$ were discussed.  Let us consider a
generic CY $p$-fold with the complex coordinates $z^i,\bar
z^i~(i=1,\dots,p)$ and the K\"ahler potential $K_0(z^i,\bar z^i)$.
The metric is given by
\bea ds^2=2\tilde g_{i\bar j} dz^i d{\bar z}^j\,,\qquad\tilde
g_{i\bar j}=\fft12\partial_i{\partial}_{\bar j}K_0= \tilde g_{\bar j
i}\,.\label{gij}
\eea 
Since it is Ricci flat, the Ricci form ${\mathcal R}^{(1,1)}$
vanishes. In fact, the Ricci-form is given by
\begin{equation}\label{Ricci}
{\mathcal R}^{(1,1)}=\i\,\tilde\partial{\bar{\tilde\partial}}
\log\sqrt{V}=0\,,
\end{equation}
where $V\equiv{\rm det}(\tilde g_{i\bar j})^2$ is the volume factor
and $\tilde \partial$ and $\bar{\tilde\partial}$ are the Dolbeault
1-form differential operators defined by
\be \tilde \partial \equiv dz^i\, \partial_{z^i}\,,\qquad
\bar{\tilde\partial} \equiv d{\bar z^i}\,\partial_{\bar z^i}\,. \ee
The equation (\ref{Ricci}) implies that $\log {V}$ is the real part
of a holomorphic function, or equivalently, $V$ can be the norm of a
holomorphic function. There is no unique choice for the complex
coordinates and we can always make a holomorphic coordinate
transformation $z^i\rightarrow z'^i=f^i(z^j)$, under which the
volume factor transforms as
\begin{equation}
V\rightarrow |T|^{-4}V\,,
\end{equation}
where
\begin{equation} T={\rm det}\left[\frac{\partial(f^1,\dots,f^p)}
{\partial (z^1,\dots,z^p)}\right]\,.
\end{equation}
The Jacobian $T$ can be any holomorphic function. Thus we can always
set $V=1$ by choosing appropriate complex coordinates. We shall do
this for later convenience.

   Let us assume that the complex vielbein for the CY $p$-fold
are $\tilde\epsilon^a, \bar{\tilde\epsilon}^a~ (a=1,\dots,p)$, then
the K\"ahler form and holomorphic $(p,0)$-form are given by
\be \tilde J^{(1,1)}=\frac{\rm
i}{2}\tilde\epsilon^a\wedge\bar{\tilde\epsilon}^a\,,\qquad
\tilde\Omega^{(p,0)}=\tilde\epsilon^1\wedge\dots\wedge
\tilde\epsilon^p\,.
\ee 
We now use this structure to construct a CY $(p+1)$-fold.  The
metric ansatz is given by
\bea\label{h}
ds^2_{2(p+1)}&=&ds^2_{2p}+h^2\;dy^2+h^{-2}(d\alpha+A)^2\cr
&=&(\delta_{a\bar
b}+G_{a\bar b})\tilde \epsilon^a\bar
{\tilde\epsilon}^b+h^2\;dy^2+h^{-2}(d\alpha+A)^2\;. \eea
The metric components are the functions of the $y,z^i,\bar z^i$
coordinates, but are independent of the coordinate $\alpha$, which
is a manifest Killing direction. Thus the metric ansatz assumes at
least one Killing direction.  The metric components appearing in the
$ds^2_{2p}$ part are defined by
\bea \tilde\partial\bar{\tilde\partial}G&=&dz^i\wedge d\bar
z^j\;\partial_{i}\partial_{\bar
j}G=\tilde\epsilon^a\wedge\bar{\tilde\epsilon}^b\;
{\tilde\epsilon}_a^i\bar{\tilde\epsilon}_b^j
\partial_{i}\partial_{\bar j}G =G_{a \bar b}\,{\tilde\epsilon}^a
\wedge\bar{\tilde\epsilon}^b\;,\label{Gabdef} \eea
where ${\tilde\epsilon}_a$ is the inverse complex vielbein. Note
that if we replace $G$ by $K_0$ in (\ref{Gabdef}), we have
$\delta_{a\bar b}$ instead of $G_{a\bar b}$.  Thus the $ds_{2p}^2$ in
(\ref{h}) is obtained by deforming the original K\"ahler potential
$K_0(z^i,\bar z^i)$ to $K_0(z^i,\bar z^i)+G(y,z^i,\bar z^i)$.

   The $ds_{2p}^2$ can be diagonalized by a local $SU(p)$ transformation
$U_a^{~b}(z^i,\bar z^i)$, namely 
\bea U\,  \Big( \delta_{a\bar b}+G_{a\bar b} \Big)\, U^{\dag}={\rm
diag} \{\lambda_1(z^i,\bar z^i) ,\dots, \lambda_p(z^i,\bar z^i)\}
\,. \eea 
We further suppose that the complex structure of the CY $p$-fold is
part of complex structure of the CY $(p+1)$-fold. This implies that
the complex vielbein of the CY $(p+1)$-fold is given by
\bea \epsilon^b=\sum_a\sqrt{\lambda_b} \,\tilde\epsilon^a
(U^{\dag})_a^{~b}\,,~~~
\epsilon^{(p+1)}=e^{\i\,{\kappa}}(h\,dy+\i\,h^{-1}(d\alpha+A))\,.
\eea
where $\kappa=\kappa(\alpha,y,z^i,\bar z^i)$ is a real function.
Correspondingly, the K\"ahler form and the $(p+1,0)$-form for the CY
$(p+1)$-fold are given by 
\bea J^{(1,1)}&=&\frac{\rm
i}{2}(\epsilon^1\wedge\bar{\epsilon}^1+\dots+
\epsilon^p\wedge\bar{\epsilon}^p+
\epsilon^{(p+1)}\wedge\bar{\epsilon}^{(p+1)})
\cr&=&\fft{\rm i}{2}(\delta_{a\bar b}+G_{a\bar b})
\tilde \epsilon^a\wedge\bar {\tilde\epsilon}^b
+dy\wedge(d\alpha+A)\;,\\
\Omega^{(p+1,0)}&=&\epsilon^1\wedge\dots\wedge\epsilon^p
\wedge\epsilon^{(p+1)}
=f\,e^{\i\,\kappa}\,\tilde\epsilon^1\wedge\dots\wedge\tilde\epsilon^p
\wedge\left(h\,dy+{\rm i}\,h^{-1}\,(d\alpha+A)\right)\,, \eea
where 
\be\label{f} f=\sqrt{\lambda_1\dots\lambda_p}= \sqrt{\det(\delta_{a\bar
b} + G_{a\bar b})}\,. \ee 

      The requirement that the metric (\ref{h}) be Calabi-Yau
becomes the requirement that the above K\"ahler form and
$(p+1,0)$-form are both closed.  Analogous to the derivation in
\cite{lpw}, we find that $dJ=0$ implies that 
\be
A=-\frac{\rm i}{4}(\tilde\partial-\bar{\tilde\partial})\partial_y G\,
+\lambda(y,z_i,\bar z_i)\,dy\,.
\ee
Note that the vanishing of $d\Omega$ implies that $\lambda$ is a
pure gauge and can be set to zero as shown in \cite{lpw}. Let us
denote $g \equiv f\,h^{-1}$ in the following. Two classes of
solutions emerges for the Calabi-Yau $(p+1)$-fold metrics,
corresponding to taking either $\kappa=\alpha$ or $\kappa=0$. They
are summarized as follows 
\bea\label{case1} \hbox{Class I:}\qquad \left\{\begin{array} {cc}
&\kappa=\alpha\,,~~~~~~~~~~~~~~~~~~~~~~~~~~~~~~~~~\\
&g=\exp\left(-\fft{1}4\partial_y G\right)\,,~~~~~~~~~~~~~~~~~\\
&\partial_y \left[\exp\left(-\fft{1}2\partial_y G\right)\right]
=2\det(\delta_{a\bar
b} + G_{a\bar b})\,;
\end{array}\right.\eea
and
\bea\label{case2} \hbox{Class II:}\qquad \left\{\begin{array} {cc}
&\kappa=0\,,~~~~~~~~~~~~~~~~~~~~~~~\\
&g=1\,,~~~~~~~~~~~~~~~~~~~~~~\\
&\partial^2_y G
=-4\det(\delta_{a\bar
b} + G_{a\bar b})\,.
\end{array}\right.
\eea

\section{Low-lying examples}

\subsection{$p=1$}

Locally, the CY1 metric is flat, namely
\bea ds^2=dz d{\bar z}\,.\eea
The K\"ahler potential is given by $K_0=z\bar z$. A proper complex
vielbein is given by $\tilde\epsilon^1=dz$ and $\bar{\tilde
\epsilon}^1 = d\bar z$.

    We first considered the $\kappa=\alpha$ case.
The system is determined solely by the following equation for $G$
\be
\partial_y \left[\exp\left(-\fft{1}2\partial_y G\right)\right]
=2(1+G_{1\bar 1} )\,.\label{basicG1p=1} \ee 
It is more convenient to use $g^2$ as the basic function in this
case. The basic equation becomes
\be\label{basicg1p=1}
\partial^2_y (g^2)+4\,\partial_z\partial_{\bar z}\log (g^2)=0\,. \ee
The corresponding CY2 metric is given by
\bea
ds^2&=&\fft{1}2\partial_y (g^2) \,dz d\bar
z+\fft{1}2g^{-2}\partial_y (g^2)\;dy^2 +\fft{2}{g^{-2}\partial_y
(g^2)}\left(d\alpha+\frac{\rm
i}{2}(\partial-\bar\partial)\,\log(g^2)\right)^2 \eea 
The general solution for (\ref{basicg1p=1}) is unknown.  A simple
way to obtain a special solution is to consider the separation of
variables. It is straightforward to show that the resulting metrics
are either ${\R}^4$ or the Eguchi-Hanson metric.

We now demonstrate that the supersymmetric limit
\cite{susy1,susy2,susy3} of the Ricci-flat Plebanski metric
\cite{pleb} is contained in (\ref{basicg1p=1}). The metric is given
by 
\be
ds^2 = \fft{({\tilde y}-{\tilde x}) d{\tilde x}^2}{4 \Delta_x} +
\fft{({\tilde y}-{\tilde x}) d{\tilde y}^2}{4 \Delta_y} +
\fft{\Delta_x}{{\tilde y}-{\tilde x}} (d\psi - {\tilde y} d\phi)^2 +
\fft{\Delta_y}{{\tilde y}-{\tilde x}} (d\psi - {\tilde x} d\phi)^2\,.
\ee 
where
\be \Delta_x = - {\tilde x}^2 + \mu\,,\qquad
\Delta_y = {\tilde y}^2 - \nu\,. \ee 
The coordinate $x$ is compact lying within
$[-\sqrt{\mu},\sqrt{\mu}]$ and the coordinate $y$ is non-compact
lying in $[\sqrt{\nu},\infty)$. The metric is singular at $\tilde
y=\tilde x$, since we have
\be {\rm Riem}^2=\fft{384(\mu-\nu)^2}{({\tilde y}-{\tilde x})^6}\,,
\ee 
but this curvature singularity can be avoided by requiring $ \mu
<\nu$.

The complex vielbein is given by 
\bea
\epsilon^1=e^{\i\psi}\left[\sqrt{ \fft{{\tilde y}-{\tilde x} }{4
\Delta_x}}d{\tilde x}+\i\,\sqrt{\fft{\Delta_x}{{\tilde y}-{\tilde
x}}} (d\psi - {\tilde y} d\phi)\right]\,,\cr
\epsilon^2=e^{\i\psi}\left[\sqrt{ \fft{{\tilde y}-{\tilde x} }{4
\Delta_y}}d{\tilde y}+\i\,\sqrt{\fft{\Delta_y}{{\tilde y}-{\tilde
x}}} (d\psi - {\tilde x} d\phi)\right]\,.\eea 
Note that the complex vielbein is defined up to a local $SU(2)$
transformation $\epsilon^a\rightarrow U^a_b\epsilon^b$. If we take
\bea U=\fft1{\sqrt{\Delta_x+\Delta_y}}\begin{pmatrix} e^{-\i\psi}& 0
  \cr 0 & e^{\i\psi}
 \end{pmatrix}\begin{pmatrix} \sqrt{\Delta_y}~~~ & -\sqrt{\Delta_x}
  \cr \sqrt{\Delta_x}~~~ & \sqrt{\Delta_y}
 \end{pmatrix}\,,\eea
then we get an expression as following
\bea \epsilon^1&=&\sqrt{ \fft{({\tilde y}-{\tilde
x})\Delta_x\Delta_y }{4 (\Delta_x+\Delta_y)}}\left(\fft{d\tilde
x}{\Delta_x}-\fft{d\tilde
y}{\Delta_y}\right)-\i\,\sqrt{\fft{({\tilde y}-{\tilde
x})\Delta_x\Delta_y}{\Delta_x+\Delta_y}}  d\phi\,,\cr
\epsilon^2&=&e^{2\i\psi}\left[\sqrt{ \fft{{\tilde y}-{\tilde x} }{4
(\Delta_x+\Delta_y)}}(d{\tilde x}+d{\tilde y}) +\i\,\sqrt{
\fft{\Delta_x+\Delta_y}{{\tilde y}-{\tilde x} }} \left(d\psi -
\fft{{\tilde x}\Delta_y+{\tilde y}\Delta_x}{\Delta_x+\Delta_y}
d\phi\right)\right]\,.\eea 
This form can be directly related to our initial ansatz, namely
\bea &&g^2=\fft14\Delta_x\Delta_y \,,~~y=\fft{{\tilde x}+{\tilde
y}}4\,,~~\alpha=2\psi\,,~~z=x_1+\i x_2\,,~~x_2=\phi\,, \cr&&
x_1=-\fft{1}{2\sqrt\mu}{\rm arctanh}\left(\fft{\tilde
x}{\sqrt\mu}\right)-\fft{1}{2\sqrt\nu}{\rm arccoth}\left(\fft{\tilde
y}{\sqrt\nu}\right)\,.\eea 
It is now straightforward to verify that $g^2$ satisfies
(\ref{basicg1p=1}).

We now consider the second case, corresponding to $\kappa=0$. The
solution is determined solely by the following basic equation for
$G$ 
\be
\partial^2_y G
+4(1+G_{1\bar 1})=0\,.\label{basicG2p=1} \ee
Note that this equation implies that $(G+2y^2)$ is a harmonic
function of the flat three-dimensional space $ds^2=dy^2 + dz d\bar
z$.  Thus $G$ can be solved completely, giving rise to the metric
\bea ds^2&=&f^2\,dzd\bar
z+f^2\,dy^2+f^{-2}\,\left[d\alpha-\frac{\rm i}{4}(dz\partial_z-d\bar
z\partial_{\bar z})\partial_y G\right]^2
\cr&=&f^2\,(dx_1^2+dx_2^2+dx_3^2)+f^{-2}\,\left[d\alpha
+\frac{1}{4}(dx_2\partial_{x_1}-dx_1\partial_{x_2})\partial_{x_3}
G\right]^2\;, \eea
where $z\equiv x_1+\i x_2$ and $x_3\equiv y$. This is exactly the
Gibbons-Hawking instanton
\be ds^2 = V^{-1} (d\alpha + A_i dx_i)^2 + V\, dx_i dx_i\,, \ee
with $V=f^2$, and the gauge fixing of $A_3=0$ by redefinition of
$\alpha$.  It is straightforward to verify that
\be\label{GH} \del_i\del_i V = 0\,,\qquad \del_i V= \epsilon_{ijk}
\del_j A_k\,. \ee
Thus the most general solution for the second case is the
Gibbons-Hawking instanton.

\subsection{$p=2$}

    The general formalism and many examples of $p=2$ was discussed
in \cite{lpw}.  Here we shall demonstrate that a non-trivial example
of cohomogeneity-two Calabi-Yau metric can be put into the form of
the general ansatz. The local metric is given by \cite{clp1,clp2}
\be ds_6^2=\fft{{\tilde x}+{\tilde y}}{4\Delta_x} d{\tilde x}^2 +
\fft{\Delta_x}{{\tilde x}+{\tilde y}} (d\tau + \fft{{\tilde
y}}{2\tilde \alpha} \,\sigma_3)^2 + \fft{{\tilde x}+{\tilde
y}}{4\Delta_y} d{\tilde y}^2 + \fft{\Delta_y}{{\tilde x}+{\tilde y}}
(d\tau -\fft{{\tilde x}}{2\tilde \alpha} \, \sigma_3)^2 +\fft{
{\tilde x}\,{\tilde y}}{4\tilde \alpha}\, (\sigma_1^2 +
\sigma_2^2)\,. \ee
where
\be
\Delta_x={\tilde x}({\tilde x}+\tilde \alpha) - \fft{2\mu}{{\tilde x}}\,,
\qquad
\Delta_y={\tilde y}(\tilde \alpha-{\tilde y}) + \fft{2\nu}{{\tilde y}}\,.
\ee
It is analogous to the $D=4$ Plebanski metric. It can be viewed
\cite{resolution} as the partial resolution of the $Y^{p,q}$ spaces
\cite{ypq}.

    The complex vielbein is given by
\bea \epsilon^1&=&e^{\i\tau}\sqrt{ \fft{ {\tilde x}\,{\tilde
y}}{4\tilde \alpha}}(\sigma_1+\i\,\sigma_2)\,,\cr
\epsilon^2&=&e^{\i\tau}\left[\sqrt{ \fft{{\tilde y}+{\tilde x}}{4
\Delta_x}}d{\tilde x}+\i\,\sqrt{\fft{\Delta_x}{{\tilde y}+{\tilde
x}}} (d\tau + \fft{{\tilde y}}{2\tilde
\alpha}\,\sigma_3)\right]\,,\cr \epsilon^3&=&e^{\i\tau}\left[\sqrt{
\fft{{\tilde y}+{\tilde x}}{4 \Delta_y}}d{\tilde
y}-\i\,\sqrt{\fft{\Delta_y}{{\tilde y}+{\tilde x}}} (d\tau -
\fft{{\tilde x}}{2\tilde \alpha}\,\sigma_3)\right]\,.\eea 

Note that the complex vielbein is defined up to a local $SU(3)$
transformation $\epsilon^a\rightarrow U^a_b\epsilon^b$. If we take
\bea U=\fft1{\sqrt{\Delta_x+\Delta_y}}\begin{pmatrix}
       e^{-\i\tau}& 0 & 0
  \cr  0 & e^{-\i\tau}& 0
  \cr 0 & 0 & e^{2\i\tau}
 \end{pmatrix}\begin{pmatrix}
 1~~~  & 0 & 0 \cr
 0~~~ & \sqrt{\Delta_y}~~~ & \sqrt{\Delta_x}
  \cr 0~~~ & -\sqrt{\Delta_x}~~~ & \sqrt{\Delta_y}
 \end{pmatrix}\,,\eea
then we get an expression as following
\bea
\epsilon^1&=&\sqrt{ \fft{ {\tilde x}\,{\tilde y}}{4\tilde
\alpha}}(\sigma_1+\i\,\sigma_2)\,,\cr \epsilon^2&=&\sqrt{
\fft{({\tilde y}+{\tilde x})\Delta_x\Delta_y }{4
(\Delta_x+\Delta_y)}}\left(\fft{d{\tilde x}}{\Delta_x}+\fft{d{\tilde
y}}{\Delta_y}\right)+\fft{\i}{2\tilde \alpha}\,\sqrt{\fft{({\tilde
y}+{\tilde x})\Delta_x\Delta_y}{\Delta_x+\Delta_y}}  \sigma_3\,,\cr
\epsilon^3&=&-e^{3\i\tau}\left[\sqrt{ \fft{{\tilde y}+{\tilde x}}{4
(\Delta_x+\Delta_y)}}(d{\tilde x}-d{\tilde y}) +\i\,\sqrt{
\fft{\Delta_x+\Delta_y}{{\tilde y}+{\tilde x}}} \left(d\tau-
\fft{{\tilde x}\Delta_y-{\tilde y}\Delta_x}{2\tilde
\alpha\,(\Delta_x+\Delta_y)} \sigma_3\right)\right]\,.\eea 
It is now straightforward to relate the solution to our ansatz.
Making the following identification 
\bea &&{d\rho}=\fft{\tilde \alpha}2\left(\fft{d{\tilde
x}}{\Delta_x}+\fft{d{\tilde
y}}{\Delta_y}\right)\,,~~~\alpha=3\tau\,,~~~~y=\fft{{\tilde
x}-{\tilde y}}6\,,\cr&& g^2=\fft{\tilde x\tilde
y\Delta_x\Delta_y}{9\tilde \alpha^3} \,, ~~~ \partial_{\rho}G+
8y=\fft{ 2\,{\tilde x}\,{\tilde y}}{\tilde \alpha}\,,\eea 
we find that $(g,G)$ satisfy 
\bea &&\partial_y g^2
=\fft{1}{8}\partial_{\rho}\left(\left(\partial_{\rho}
G+8y\right)^2\right)\,, \cr && \partial_y(\partial_{\rho}G+
8y)+2\partial_\rho\log g^2-8=0\,. \eea 
It is straightforward now to verify that the basic equation
(\ref{case1}) is satisfied.  Thus the resolved $Y^{p,q}$ cone is
indeed a class I solution of our basic construction, although it is
hard to obtain directly by solving the basic equation.

\subsection{$p=3$}

With the increasing value of $p$, the basic equation becomes more
and become non-linear and difficult to solve.  Here we shall again
only demonstrate that a non-trivial example of previously-known
Calabi-Yau metric can indeed put into the form of the general
ansatz.

The metric is cohomogeneity-2 and given by \cite{clp1,clp2}
\bea ds^2&=&\fft{{\tilde x}+{\tilde y}}{4\Delta_x} d{\tilde x}^2 +
\fft{\Delta_x}{{\tilde x}+{\tilde y}} \left(d\tau + \fft{{\tilde
y}}{2\tilde \alpha} \,(d\beta + \gamma^2 \sigma_3)\right)^2 \cr&&+
\fft{{\tilde x}+{\tilde y}}{4\Delta_y} d{\tilde y}^2 +
\fft{\Delta_y}{{\tilde x}+{\tilde y}} \left(d\tau -\fft{{\tilde
x}}{2\tilde \alpha} \, (d\beta + \gamma^2 \sigma_3)\right)^2
\cr&&+\fft{ {\tilde x}\,{\tilde y}}{\tilde \alpha}\,
\Big(\fft{d\gamma^2}{V_0} + \ft14 V_0 \gamma^2 \sigma_3^2 + \ft14
\gamma^2(\sigma_1^2 + \sigma_2^2)\Big)\,. \eea
where
\be \Delta_x={\tilde x}({\tilde x}+\tilde \alpha) -
\fft{2\mu}{{\tilde x}^2}\,,\qquad \Delta_y={\tilde y}(\tilde
\alpha-{\tilde y}) + \fft{2\nu}{{\tilde y}^2}\,. \ee
The complex vielbein is given by 
\bea \epsilon^1&=&e^{\i\tau+\i\ft{3}{2}\beta}\sqrt{ \fft{ {\tilde
x}\,{\tilde y}}{\tilde \alpha}}\left(\fft{d\gamma}{\sqrt{V_0}} +
\ft{\rm i}2\gamma\, {\sqrt{V_0}}\,\sigma_3\right)\,,\cr
\epsilon^2&=&e^{\i\tau}\,\gamma\,\sqrt{ \fft{ {\tilde x}\,{\tilde
y}}{4\tilde \alpha}}(\sigma_1+\i\,\sigma_2)\,,\cr
\epsilon^3&=&e^{\i\tau}\left[\sqrt{ \fft{{\tilde y}+{\tilde x}}{4
\Delta_x}}d{\tilde x}+\i\,\sqrt{\fft{\Delta_x}{{\tilde y}+{\tilde
x}}} \left(d\tau + \fft{{\tilde y}}{2\tilde \alpha}\,(d\beta +
\gamma^2 \sigma_3)\right)\right]\,,\cr
\epsilon^4&=&e^{\i\tau}\left[\sqrt{ \fft{{\tilde y}+{\tilde x}}{4
\Delta_y}}d{\tilde y}-\i\,\sqrt{\fft{\Delta_y}{{\tilde y}+{\tilde
x}}} \left(d\tau -\fft{{\tilde x}}{2\tilde \alpha}\, (d\beta +
\gamma^2 \sigma_3)\right)\right]\,.\eea 
Making the following $SU(4)$ transformation 
\bea U=\fft1{\sqrt{\Delta_x+\Delta_y}}\begin{pmatrix}
       e^{-\i\tau} & 0 & 0 & 0
  \cr  0 & e^{-\i\tau} & 0 & 0
  \cr  0 & 0 & e^{-\i\tau} & 0
  \cr  0 & 0 & 0 & e^{3\i\tau}
 \end{pmatrix}\begin{pmatrix}
 1~~~  & 0~~~ & 0 & 0 \cr
 0~~~ & 1~~~  & 0 & 0 \cr
0~~~ & 0~~~ & \sqrt{\Delta_y}~~~ & \sqrt{\Delta_x}
  \cr 0~~~ &0~~~ & -\sqrt{\Delta_x}~~~ & \sqrt{\Delta_y}
 \end{pmatrix}\,,\eea
we get an expression as following 
\bea \epsilon^1&=&e^{\i\ft{3}{2}\beta}\sqrt{ \fft{ {\tilde
x}\,{\tilde y}}{\tilde \alpha}}\left(\fft{d\gamma}{\sqrt{V_0}} +
\ft{\rm i}2\gamma\, {\sqrt{V_0}}\,\sigma_3\right)\,,\cr
\epsilon^2&=&\sqrt{ \fft{ {\tilde x}\,{\tilde y}}{4\tilde
\alpha}}(\sigma_1+\i\,\sigma_2)\,,\cr \epsilon^3&=&\sqrt{
\fft{({\tilde y}+{\tilde x})\Delta_x\Delta_y }{4
(\Delta_x+\Delta_y)}}\left(\fft{d{\tilde x}}{\Delta_x}+\fft{d{\tilde
y}}{\Delta_y}\right)+\fft{\i}{2\tilde \alpha}\,\sqrt{\fft{({\tilde
y}+{\tilde x})\Delta_x\Delta_y}{\Delta_x+\Delta_y}}  (d\beta +
\gamma^2 \sigma_3)\,,\\
 \epsilon^4&=&-e^{4\i\tau}\left[\sqrt{
\fft{{\tilde y}+{\tilde x}}{4 (\Delta_x+\Delta_y)}}(d{\tilde
x}-d{\tilde y}) +\i\,\sqrt{ \fft{\Delta_x+\Delta_y}{{\tilde
y}+{\tilde x}}} \left(d\tau- \fft{{\tilde x}\Delta_y-{\tilde
y}\Delta_x}{2\tilde \alpha\,(\Delta_x+\Delta_y)} (d\beta + \gamma^2
\sigma_3)\right)\right]\,.\nn\eea 
To relate to our original anatz, we make the following
identification
\bea &&{d\rho}=\fft{\tilde \alpha}2\left(\fft{d{\tilde
x}}{\Delta_x}+\fft{d{\tilde
y}}{\Delta_y}\right)\,,~~~\alpha=4\tau\,,~~~~y=\fft{{\tilde
x}-{\tilde y}}8\,,\cr&& g^2=\fft{\tilde x^2\tilde
y^2\Delta_x\Delta_y}{16\,\tilde \alpha^4} \,, ~~~
\partial_{\rho}G+ 12y=\fft{ 2\,{\tilde x}\,{\tilde y}}{\tilde
\alpha}\,.\eea 
We also find the following relations
\bea &&\partial_y g^2
=\fft{1}{24}\partial_{\rho}\left(\left(\partial_{\rho}
G+12y\right)^3\right)\,, \cr && \partial_y(\partial_{\rho} G+
12y)+2\partial_\rho\log g^2-12=0\,.\eea 
Then it is easy to verify that $g$ and $G$ satisfy the basic
equation (\ref{case1}). Thus this non-trivial CY3 metric is indeed a
class I solution of our basic construction, although it is hard to
obtain directly by solving the basic equation.

\section{The holomorphic linearization}

    The main obstacle of solving the basic equations (\ref{case1}) and
(\ref{case2}) is their non-linearity.  As shown in section 3.1, for
$\kappa=0$ and $p=1$, the equation becomes linear and it can be
solved completely, giving rise to the Gibbons-Hawking instanton. We
are now looking for a subset of CY $p$-folds such that the non-linear
terms of the resulting basic equations for $(p+1)$-folds vanish.  We
shall focus our attention on $p=2$, for which, when $\kappa=0$, the
basic equation is given by 
\be
\partial^2_y G
+4(1+G_{1\bar 1}+G_{2\bar 2}+G_{1\bar 1}G_{2\bar 2}-G_{1\bar
2}G_{2\bar 1})=0\,.\ee
If we have
\be G_{1\bar 1}G_{2\bar 2}-G_{1\bar
2}G_{2\bar 1}={\rm constant}\,,\ee
we shall be left with a linear equation just like the one in the
$p=1$ case.

An immediate example is the holomorphic linearization discussed in
\cite{fay}. (See also \cite{sant}.) In this case, the function $G$
is of the form $G=G(\omega,\bar\omega,y)$ where
$\omega=\omega(z_1,z_2)$ is an arbitrary holomorphic function. Then
it is easy to show that $G_{1\bar 1}G_{2\bar 2}-G_{1\bar2}G_{2\bar
1}=0$ by noting that $G_{i\bar j}=\omega_i\bar \omega_{\bar
j}G_{\omega\bar\omega}$, where $\omega_i=\partial_{z_i} \omega$. The
basic equation now takes the form 
\bea
\partial^2_y G&=& -4(1+G_{1\bar 1}+G_{2\bar 2})=-4(1+
\sum_a{\tilde\epsilon}_a^i\bar{\tilde\epsilon}_a^j
\partial_{i}\partial_{\bar j}G)=-4(1+\tilde g^{i\bar j}
\partial_{i}\partial_{\bar j}G)
\cr&=&-4(1+\triangle_4G)=-4(1+\tilde g^{i\bar j}\omega_i\bar
\omega_{\bar j}G_{\omega\bar\omega})\,,\eea
where $\triangle_4$ is the Laplacian on the four dimensional base
and we have used $G=G(\omega,\bar\omega,y)$ at the last step. We can
always take a new complex coordinate system $\{\tilde z_1,\tilde
z_2\}$ where $\tilde z_1=\omega$, and then drop off the tilde.
Therefore, the holomorphic linearization is equivalent to take
$G=G(z_1,\bar z_1,y)$ in our general construction. In this
coordinate system, the basic equation takes the form 
\be\label{basiclinear}
\partial^2_y G=-4(1+\tilde g^{1\bar 1}\partial_{1}
\partial_{\bar 1}G)\,.\ee
The consequence is that the base CY2 space is now restricted so that
we have $\tilde g^{1\bar 1}=\tilde g^{1\bar 1}(z_1,\bar z_1)$. As
shown in section 2, we can impose $V=1$ without loss of generality,
for which, we have $\tilde g_{2\bar2}=\tilde g^{1\bar 1}$. It
follows from (\ref{gij}) that the K\"ahler potential for the
four-dimensional base must take the following form 
\bea K_0=K^{(1)}_0(z_1,\bar z_1)+z_2\bar z_2\,K^{(2)}_0(z_1,\bar
z_1)+\,K^{(3)}_0(z_1,\bar z_1,z_2)+{\bar K^{(3)}_0(\bar z_1,
z_1,\bar z_2)}\,.\eea 
The condition $V=1$ becomes
\bea\label{V=1} &&(\partial_{1}\partial_{\bar
1}K^{(1)}_0+\partial_{1}\partial_{\bar
1}K^{(3)}_0+\partial_{1}\partial_{\bar 1}\bar K^{(3)}_0+z_2\bar
z_2\,\partial_{1}\partial_{\bar 1}K^{(2)}_0)K^{(2)}_0
\cr&&-(\partial_{1}\partial_{\bar 2}\bar
K^{(3)}_0+z_2\,\partial_{1}K^{(2)}_0) (\partial_{\bar
1}\partial_{2}K^{(3)}_0+\bar z_2\,\partial_{\bar 1}\bar
K^{(2)}_0)=1\,.\eea 
Performing $\partial_{2}\partial_{\bar 2}$ on the both sides, we
have
\bea
K^{(2)}_0\partial_{1}\partial_{\bar 1}K^{(2)}_0
-\partial_{1}\partial^2_{\bar 2}\bar K^{(3)}_0\partial_{\bar 1}
\partial^2_{2}K^{(3)}_0
-\partial_{\bar 1}\bar K^{(2)}_0\partial_{1}K^{(2)}_0=0\,.\eea
It implies that
\be K^{(3)}_0(z_1,\bar z_1,z_2)=z_2^2\,k_2(z_1,\bar z_1)+z_2\,
k_1(z_1,\bar z_1)\,.\ee 
Substituting this into (\ref{V=1}), we obtain a polynomial of $z_2$
and $\bar z_2$ that vanishes.  The vanishing of the coefficients of
all the powers of $z_2$ and $\bar z_2$ implies that 
\bea \label{V=1.1}
K^{(2)}_0\,\partial_{1}\partial_{\bar
1}K^{(1)}_0-\partial_{1}\bar k_1 \,\partial_{\bar 1}k_1&=&1\,,
\\\label{V=1.2} K^{(2)}_0\partial_{1}\partial_{\bar 1}k_1-\partial_{1}
K^{(2)}_0\partial_{\bar 1}k_1
&=&2\,\partial_{\bar 1} k_2\,\partial_{1}\bar k_1\,,
\\ \label{V=1.3}
K^{(2)}_0\partial_{1}\partial_{\bar 1}K^{(2)}_0-\partial_{1}K^{(2)}_0
\partial_{\bar 1}\bar K^{(2)}_0
&=&4\,\partial_{1}\bar k_2\,\partial_{\bar 1} k_2\,,
\\ \label{V=1.4} K^{(2)}_0\partial_{1}\partial_{\bar 1}k_2-2
\partial_{1}K^{(2)}_0\partial_{\bar 1}k_2&=&0\,.\eea

   One class of solutions is that $k_2\equiv0$, for which,
the equation (\ref{V=1.3}) can be solved as follows 
\be K^{(2)}_0=F(z_1)\bar F(\bar z_1)\,.\ee
The equation (\ref{V=1.2}) then implies that $\partial_{\bar
1}k_1/K_0^{(2)}$ is an anti-holomorphic function. It can be shown
that up to a holomorphic gauge transformation for the K\"ahler
potential, we can also write $k_1/K_0^{(2)}=\bar h(\bar z_1)$. By
using the holomorphic coordinate transformation $z_1\rightarrow z_1$
and $z_2\rightarrow z_2-h(z_1)$, we can further set $K^{(3)}_0=0$
while preserving our ansatz. By using (\ref{V=1.1}), we find 
\bea K_0=H(z_1)\bar H(\bar z_1)+z_2\bar z_2\,F(z_1)\bar F(\bar
z_1)\,,~~~~H(z_1)=\int \fft1{F(z_1)} dz_1.\eea 
Note that the condition $G=G(z_1,\bar z_1,y)$ is equivalent to the
condition $G=G(H,\bar H,y)$. Thus we can always perform the
transformation $\tilde z_1=H(z_1)$ and $\tilde z_2=z_2F(z_1)$
without violating the ansatz\footnote{The two examples discussed in
Sec7.1 \& 7.2 of \cite{fay} are indeed the same one. The solution in
section 7.2 of \cite{fay} corresponds to take $F=\sqrt {z_1}$ in our
above discussion. }. Therefore, the base space is just the ${\R}^4$
and the corresponding CY3-fold is simply the direct product of the
Gibbons-Hawking instanton and ${\R}^2$. This case was discussed in
detail in \cite{fay,sant}.

We shall now focus our attention on the new case with non-vanishing
$k_2$. The solution to (\ref{V=1.4}) is given by 
\be \partial_{\bar 1} k_2=\bar h_2(\bar z_1)(K^{(2)}_0)^2\,. \ee 
By the holomorphic coordinate transformation $\tilde z_1(z_1)=\int
h_2(z_1)dz_1$ we can absorb $h_2$. It is necessary to make a
coordinate transformation $\tilde z_2=z_2/h_2(z_1)$ in order to
preserve $V=1$. However this has no effect on the equations
(\ref{V=1.1})-(\ref{V=1.4}).  Thus this transformation is equivalent
to set $h_2=1$. Then (\ref{V=1.3}) and (\ref{V=1.2}) become 
\bea\label{VII}
\partial_{1}\partial_{\bar 1}\log{K^{(2)}_0}
&=&4(K^{(2)}_0)^2\,,~~~\partial_{1}
\left(\fft{\partial_{\bar 1}k_1}{K^{(2)}_0}\right)
=2\partial_{1}\bar k_1\,.\eea
Note that $K_0^{(1)}$ can then be determined by (\ref{V=1.1}).

The first equation of (\ref{VII}) is precisely the two-dimensional
Liouville equation. Once the $K_0^{(2)}$ is solved, the remaining
functions follow straightforwardly.  Thus, the corresponding CY3
space is governed in essence by the solutions of the Liouville
equation. The general solutions to the Liouville equation are not
known.  In \cite{liouref}, many special solutions were given.  Here,
we examine two examples in detail.

The first example of the special solutions of the Liouville equation
is given by \cite{liouref}
\be \label{VII.1} (K^{(2)}_0)^2=\fft{F'(z_1)\bar F'(\bar
z_1)}{4\,(F(z_1)+\bar F(\bar z_1))^2}\,.\ee 
(Note that in \cite{liouref} the solution is more general in that
the $\bar F$ is replaced by an unrelated anti-holomorphic function
to $F$.  Here we chose it to be $\bar F$ so that $K_0^{(2)}$ is
real.) Consequently we have 
 \be k_2=-\fft{F'}{4\,(F+\bar F)}\,.\ee
Up to a gauge transformation of the K\"ahler potential, the second
equation of (\ref{VII}) implies 
\bea\label{k1}
\partial_{\bar 1}k_1
=2{K^{(2)}_0}\bar k_1=\fft{\big(F'\bar F'\big)^{\fft12}}{F
+\bar F}\bar k_1\,.\eea
It can be shown further that we can always fix $k_1=0$ by choosing
proper complex coordinates. Thus we have 
\bea
&&\partial_{1}\partial_{\bar 1}K^{(1)}_0=\fft{2\,(F+\bar
F)}{\big(F'\bar F'\big)^{\fft12}}\,,~~~~ \cr&&z_2\bar
z_2\,K^{(2)}_0(z_1,\bar z_1)+K^{(3)}_0(z_1,\bar z_1,z_2)+\bar
K^{(3)}_0(\bar z_1, z_1,\bar z_2) =\fft{z_2\bar z_2\big(F'\bar
F'\big)^{\fft12}}{2\,(F+\bar F)}-\fft{z_2^2F'+\bar z_2^2\bar
F'}{4\,(F+\bar F)} \,.~~~~~~\eea 
Making the coordinate transformation $\tilde z_1(z_1)=\int
F'^{-\fft12}dz_1$ and $\tilde z_2=z_2F'^{\fft12}$, and then dropping
off the tildes, the K\"aher potential becomes 
\be K_{0}=2\,(H_1+\bar H_1)(z_1+\bar z_1)-\fft{(z_2-\bar
z_2)^2}{4\,(F+\bar F)}\,,\qquad H_1(z_1)=\int Fdz_1\,.
\label{example1k0} \ee 
The corresponding CY2 base is given by
\bea\label{sII1} ds^2&=&\Big[2(F+\bar F)-\fft{(z_2-\bar z_2)^2F'\bar
F'}{2(F +\bar F)^{3}}\Big]dz_1d\bar z_1 +\fft{1}{2(F+\bar
F)}dz_2d\bar z_2 \cr&&-\fft{(z_2-\bar z_2)F'}{2(F+\bar
F)^{2}}dz_1d\bar z_2 +\fft{(z_2-\bar z_2)\bar F'}{2(F+\bar
F)^{2}}dz_2d\bar z_1 \,.
\eea 
There is a curvature singularity at $F+\bar F=0$.
The proper complex vielbein is given by
\bea
\tilde\epsilon^1&=&\sqrt{2}\,(F+\bar F)^{\fft12}dz_1\,,
\cr\tilde\epsilon^2&=&-\fft{(z_2-\bar z_2)F'}{\sqrt{2}\,(F
+\bar F)^{\fft32}}dz_1
+\fft{1}{\sqrt{2}\,(F+\bar F)^{\fft12}}dz_2\,.\eea
The metric (\ref{sII1}) is in fact a special case of the
Gibbons-Hawking solutions.  This can be seen from the fact that the
metric has a Killing direction $(z_2 + \tilde z_2)$.  To see this in
detail, let $F(z_1)=u+\i\, v$, $z_2=\alpha+\i \,w$ and $\tilde
y=\fft{w}{4 u}$. Then we find 
\bea
\tilde\epsilon^1&=&2\,u^{\fft12}\,(dx_1+\i\,dx_2)\,,
\cr\tilde\epsilon^2&=&-\fft{2\,\i\, w }{\sqrt{2}\,(2u)^{\fft32}}
(du+\i\, dv)
+\fft{1}{\sqrt{2}\,(2u)^{\fft12}}(d\alpha+\i \,dw)
\cr&=&\fft{1}{2\,u^{\fft12}}(d\alpha+\fft1{2u} \,dv)
+\i\,2\,u^{\fft12}d\tilde y\,\,.\eea
The metric corresponds to the following solution of
(\ref{basicG2p=1}) for the function $\tilde G$ 
\be\tilde G=-8\,\tilde y\,u^{\fft12}-z_1\bar z_1+ 2\int
u^{\fft12}\,dz_1d\bar z_1\,.\ee

Having obtained the CY2 base, we now only need to solve the
function $G$ in order to obtain the corresponding CY3. By
construction, the basic equation for $G$ is linear and given by 
\be\label{basicII.1}
\partial^2_y G+4+2{(F+\bar F)}^{-1}\partial_{1}\partial_{\bar 1}G=0
\,.\ee
This means that $(G+2y^2)$ is the harmonic function on the space 
\be ds_3^2=dy^2+2(F+\bar F)dz_1d\bar z_1\,.\ee 
Note that the three-space is a direct product of $\R$ associated
with the coordinate $y$ and a two-dimensional K\"ahler space
associated with the coordiantes $(z_1,\bar z_1)$. The K\"ahler
potential is given by $K_0^{(1)}$, {\it i.e.}~it is given by the
first term of $K_0$ given in (\ref{example1k0}). Thus the nature of
the CY3 is determined by the property of the two-dimensional
K\"ahler space, and in particular by the holomorphic function $F$.
The CY3 metric is given by 
\bea\label{GH111} ds^2 &=& f^2 ds_3^2+f^{-2}(d\alpha + A)^2
-\fft{(z_2-\bar z_2)^2F'\bar F'}{2(F +\bar F)^{3}}dz_1d\bar z_1\cr&&
+\fft{1}{2(F+\bar F)}dz_2d\bar z_2 -\fft{(z_2-\bar z_2)F'}{2(F+\bar
F)^{2}}dz_1d\bar z_2 +\fft{(z_2-\bar z_2)\bar F'}{2(F+\bar
F)^{2}}dz_2d\bar z_1\,,\cr f^2&=&1+\fft1{2(F+\bar
F)}\partial_1\partial_{\bar 1}G=-\fft14\partial^2_yG\,, \qquad
A=\frac{1}{4}(dx_2\partial_{x_1} -dx_1\partial_{x_2})\partial_{x_3}
G\,, \eea 
where $z_1=x_1 + {\rm i} x_2$ and $y=x_3$. It is straightforward to
verify that we have $\triangle_3 f^2=0$ and $d(f^2)=*_3dA$, where
$\triangle_3$ and the Hodge dual $*_3$ are defined with respect to
the metric $ds_3^2$.  We can rewrite the new CY3 metric in the
following form
\bea\label{GH112} ds^2 &=& V^{-1} (d\alpha + A_i dx_i)^2 + V\,
ds_3^2 -\fft{(z_2-\bar z_2)^2F'\bar F'}{2(F +\bar F)^{3}}dz_1d\bar
z_1\cr && +\fft{1}{2(F+\bar F)}dz_2d\bar z_2 -\fft{(z_2-\bar
z_2)F'}{2(F+\bar F)^{2}}dz_1d\bar z_2 +\fft{(z_2-\bar z_2)\bar
F'}{2(F+\bar F)^{2}}dz_2d\bar z_1\,, \cr \triangle_3 V &=&
0\,,\qquad dV= *_3 d A\,.
\eea 
Note that in (\ref{GH111}) the component $A_3$ is fixed to be zero
by appropriate shifting of the $\alpha$ coordinate.

    The metric (\ref{GH112}) is the six-dimensional analogue of
Gibbons-Hawking instanton.  It can also be viewed as an $\R^2$
bundle over a four-space ${\cal X}_4$, where the bundle coordinates
are $(z_2,\bar z_2)$ and the base-space coordinates are $(\alpha, y,
z_1, \bar z_1)$. The base space ${\cal X}_4$ depends on two
functions. One is an arbitrary holomorphic function $F$ of
coordinate $z_1$. The other is the harmonic function $V$ on a
generically non-flat three-dimensional space with the metric
$ds_3^2$.  When $F$ is a constant, the base metric is the
Gibbons-Hawking instanton, and corresponding CY3 is the instanton
appended by an $\R^2$.

For non-constant $F$, as a simple demonstrative example, we may take
$F=z_1$, the analysis of $z_1+\bar z_1$=constant surface suggests that this
CY2 metric describes the $a\rightarrow\infty$ limit of the
Eguchi-Hanson space. The basic equation (\ref{basiclinear}) becomes
\be\label{basiclinear2}
\partial^2_y G+4+2{(z_1+\bar z_1)}^{-1}\partial_{1}
\partial_{\bar 1}G=0\,.\ee
The general solution is given by
\bea
&&G=-2y^2+\sum_{p_0,p_2}e^{\i (p_0y+ p_2x_2)}
\Big(\lambda^{(1)}_{p_0,p_2}
{\rm Ai}(u)+\lambda^{(2)}_{p_0,p_2}{\rm Bi}(u)\Big)\,,\cr&&
u=p_0^{\fft23}\,x_1+p_2^2\,p_0^{-\fft43}\,,~~~z_1=x_1+\i x_2\,,\eea
where ${\rm Ai}(u)$ and  ${\rm Bi}(u)$ are the Airy functions.

Another example of the special solutions to the Liouville equation
is given by \cite{liouref} 
\be \label{VII.2} (K^{(2)}_0)^2=\fft{F'(z_1)\bar F'(\bar
z_1)}{4(1-F(z_1) \bar F(\bar z_1))^{2}}\,.\ee 
Then we have
\be k_2=\fft{F'\bar F}{4\,(1-F\bar F)}\,.\ee Up to a gauge
transformation of the K\"ahler potential, the second equation of
(\ref{VII}) implies 
\bea\label{k1_II.2}
\partial_{\bar 1}k_1
=2{K^{(2)}_0}\bar k_1=\fft{\big(F'\bar F'\big)^{\fft12}}{1
-F\bar F}\bar k_1\,.\eea
We take a simple solution of $k_1=0$, it follows that we have
\bea &&\partial_{1}\partial_{\bar 1}K^{(1)}_0=\fft{2\,(1-F\bar
F)}{\big(F'\bar F'\big)^{\fft12}}\,,~~~~\\&&z_2\bar
z_2\,K^{(2)}_0(z_1,\bar z_1)+K^{(3)}_0(z_1,\bar z_1,z_2)+\bar
K^{(3)}_0(\bar z_1, z_1,\bar z_2) =\fft{z_2\bar z_2\big(F'\bar
F'\big)^{\fft12}}{2\,(1-F\bar F)}+\fft{z_2^2F'\bar F+\bar z_2^2F\bar
F'}{4\,(1-F\bar F)} \,.\nn\eea 
After the coordinate transformation $\tilde z_1(z_1)=\int
F'^{-\fft12}dz_1$ and $\tilde z_2=z_2F'^{\fft12}$ and dropping off
the tildes afterwards, the K\"ahler potential becomes 
\be K_{0}=2\,(z_1\bar z_1-H_1\bar H_1)+\fft{z_2^2\bar F+\bar
z_2^2F+2z_2\bar z_2}{4\,(1-F\bar F)}\,,\qquad H_1(z_1)=\int
Fdz_1\,.\label{example2k0}
\ee 
The corresponding CY2 base is given by
\bea\label{sII2} ds^2&=&\Big[2(1-F\bar F)+\fft{z_2^2\bar F+\bar
z_2^2F+z_2\bar z_2(1+F\bar F)}{2(1-F\bar F)^{3}}F'\bar
F'\Big]dz_1d\bar z_1 +\fft{1}{2(1-F\bar F)}dz_2d\bar z_2
\cr&&+\fft{z_2\bar F+\bar z_2}{2(1-F\bar F)^{2}}F' dz_1d\bar z_2
+\fft{z_2+\bar z_2F}{2(1-F\bar F)^{2}}\bar F'dz_2d\bar z_1
\,.~~~\eea 
There is a curvature singularity at $F\bar F=1$.
The proper complex vielbein is given by
\bea
\tilde\epsilon^1&=&\sqrt{2}\,(1-F\bar F)^{\fft12}dz_1\,,
\cr\tilde\epsilon^2&=&\fft{(z_2 \bar F+\bar z_2)F'}{\sqrt{2}\,
(1-F\bar F)^{\fft32}}dz_1
+\fft{1}{\sqrt{2}\,(1-F\bar F)^{\fft12}}dz_2\,.\eea
The metric (\ref{sII2}) does not have a Killing direction and hence
lies outside the construction of $p=1$ case discussed in section 3.
To obtain the corresponding CY3 solution, we need to solve the basic
equation 
\be\label{basiclinear3}
\partial^2_y G+4+2{(1-F\bar F)}^{-1}\partial_{1}
\partial_{\bar 1}G=0\,.\ee
It means that $(G+2y^2)$ is the harmonic function on the space
\be ds_3^2=dy^2+2(1-F\bar F)dz_1d\bar z_1\,.\ee
Again this metric is a direct product of $\R$ and a two-dimensional
K\"ahler space whose K\"ahler potential is given by the first term
of $K_0$ given by (\ref{example2k0}). The corresponding CY3 metric
is given by 
\bea\label{GH121} ds^2 &=&f^2
ds_3^2+f^{-2}\,\left[d\alpha+\frac{1}{4}(dx_2\partial_{x_1}-dx_1
\partial_{x_2})\partial_{x_3}
G\right]^2 \cr&&+\fft{z_2^2\bar F+\bar z_2^2F+z_2\bar z_2(1+F\bar
F)}{2(1-F\bar F)^{3}}F'\bar F'dz_1d\bar z_1 +\fft{1}{2(1-F\bar
F)}dz_2d\bar z_2 \cr&&+\fft{z_2\bar F+\bar z_2}{2(1-F\bar F)^{2}}F'
dz_1d\bar z_2 +\fft{z_2+\bar z_2F}{2(1-F\bar F)^{2}}\bar F'dz_2d\bar
z_1 \,,\cr f^2&=&1+\fft1{2(1-F\bar F)}\partial_1\partial_{\bar
1}G=-\fft14\partial^2_yG\,.\eea 
As in the previous example, we can rewrite the metric in the
following form
\bea\label{GH122} ds^2 &=& V^{-1} (d\alpha + A_i dx_i)^2 + V\,
ds_3^2 \cr&&+\fft{z_2^2\bar F+\bar z_2^2F+z_2\bar z_2(1+F\bar
F)}{2(1-F\bar F)^{3}}F'\bar F'dz_1d\bar z_1 +\fft{1}{2(1-F\bar
F)}dz_2d\bar z_2 \cr&&+\fft{z_2\bar F+\bar z_2}{2(1-F\bar F)^{2}}F'
dz_1d\bar z_2 +\fft{z_2+\bar z_2F}{2(1-F\bar F)^{2}}\bar F'dz_2d\bar
z_1\,, \cr \triangle_3 V &=& 0\,,\qquad dV= *_3 d A\,. \eea
Analogous to the previous example, the metric is a generalization of
the Gibbons-Hawking instanton.  The metrics depends on two functions
$F$ and $V$.  $F$ is an arbitrary holomorphic function and $V$ is
the harmonic function in $ds_3^2$.  For constant $F$, the metric is
a direct product of Gibbons-Hawking instanton and $\R^2$.  For
non-constant $F$, as a demonstrative example, we consider $F=z_1$.
Then we have 
\bea &&G=-2y^2+\sum_{p_0,p_2}e^{\i (p_0y+
p_2\theta)}2^{\fft{p_2+1}2}r^{p_2}e^{-\fft{\i p_0
r^2}{2\sqrt{2}}}\Big(\lambda^{(1)}_{p_0,p_2}W_1(p_0,p_2,r)
+\lambda^{(2)}_{p_0,p_2}W_2(p_0,p_1,r)\Big)\,,\cr&&
W_1(p_1,p_2,r)=U\left(4(p_2+1)-\i\sqrt2p_0,p_2+1,\fft{\i p_0
r^2}{\sqrt{2}}\right)\,,\cr&&
W_2(p_1,p_2,r)=L_{-4(p_2+1)+\i\sqrt2p_0}^{p_2}\left(\fft{\i p_0
r^2}{\sqrt{2}}\right)\,,~~~z_1=r\,e^{\i\theta}\,,\eea 
where $U$ is the confluent hypergeometric function and $L$ is the
generalized Laguerre polynomial.

    It turns out that the above two examples can be unified to give
rise to a more general solution.  The metric becomes more
complicated, but the essential properties remain the same.  We
present the detail discussion in the appendix.

There are still some other special special solutions of the
Liouville equation which are known as functional separable
solutions. The explicit forms are \cite{liouref}
\bea (K^{(2)}_0)^2&=&\fft{1}{(c_1\,e^{x_1}\pm4\cos x_2)^2}\,,\\
(K^{(2)}_0)^2&=&\fft{c_2^2-c_1^2}{16(c_1\,\cosh{x_1}+c_2\,\sin
x_2)^2}\,,\\
(K^{(2)}_0)^2&=&\fft{c_2^2+c_1^2}{16(c_1\,\sinh{x_1}+c_2\,\cos
x_2)^2}\,. \eea
where $c_1$ and $c_2$ are arbitrary constants and $x_1 + {\rm i} x_2
= z_1$. These Liouville solutions will lead to different CY2 bases
and CY3 metrics.  We shall not analyze them further.

\section{Conclusions}

In this paper, we establish the general formalism for constructing
metrics of Calabi-Yau $(p+1)$-folds in terms of that of a $p$-fold
by adding a complex-line bundle. We present a few explicit low-lying
examples.  The metrics are determined by the basic equations for the
function $G$, given by (\ref{case1}) or (\ref{case2}). The
obstacle to solve these equations is the higher non-linearity for
higher $p$.  For $p=1$, the equation for $G$ in (\ref{case2})
becomes linear and the resulting solution is the Gibbons-Hawking
instanton.

     For $p=2$, we consider holomorphic linearization and focus on
subset of solutions where the non-linear terms in (\ref{case2})
vanish.  This restricts the possible CY2 base spaces, which turn out
to be governed by a two-dimensional Liouville equation.  The
K\"ahler potential for the CY3 is then given by a harmonic function
in certain three-dimensional non-flat spaces, depending on the
specific solution of the Liouville equation.  We provide detail
analysis for two such special solutions of the Liouville equation.
In both of these examples, the metric can be viewed as a
complex-line bundle over a four-dimensional space ${\cal X}_4$ which
can be viewed as an $U(1)$ bundle over a three dimensional space
that is a direct product of $\R$ and a two-dimensional K\"ahler
space. The two-dimensional K\"ahler space is determined by an
arbitrary holomorphic function. When the holomorphic function is a
pure constant, the three-space becomes flat and ${\cal X}_4$ becomes
precisely the Gibbons-Hawking instanton.  Alternatively, by our
construction, these solutions are complex-line bundle over CY2
bases.  In particular, the CY2 metrics (\ref{sII2}) and
(\ref{generalcy2}) we obtained are highly non-trivial in that they
do not have Killing direction and lie outside our construction for
$p=1$.

  It is of great interest to investigate further our new solutions for
generic holomorphic functions and examine the global structure of
the resulting CY2 bases and corresponding CY3 metrics. We shall
study this in a future publication.

    We expect the linearization procedure can be generalized further
to higher-dimensions and give rise to the higher-dimensional
analogue of Gibbons-Hawking instanton.

\appendix

\section*{Appendix: more general CY2 and CY3 metrics}

In section 4, we demonstrate that for some special choice of the CY2
base space, the function $G$ is governed by a linear
equation. The key property of the base space is that it is
determined by the solutions of the two-dimensional Liouville
equation, namely 
\be
\partial_{1}\partial_{\bar 1}\log{K^{(2)}_0}=4(K^{(2)}_0)^2\,,\ee
Consulting the mathematics reference book \cite{liouref}, we
presented two special solutions and discuss the resulting CY2 and
CY3 metrics in details. In this appendix, we show that these two
solutions can in fact be unified into one more general solution; it
is given by
\bea  (K^{(2)}_0)^2\!\!&=&\!\!\fft{F_1'(z_1)\bar F_2'(\bar
z_1)}{4}\left(\fft{a}{(1-a\,F_1(z_1)\bar F_2(\bar
z_1))^2}\!+\!\fft{1-a\,b^2}{(b+F_1(z_1)+\bar F_2(\bar
z_1)+a\,b\,F_1(z_1)\bar F_2(\bar z_1))^2}\right)\cr&=&\fft{F_1'\bar
F_2'}{4} \fft{{(a\,F_1^2+2\,a\,b\, F_1+1)(a\,{\bar F_2}^2+2\,a\,b\,
{\bar F_2}+1)}} {{(1-a\,F_1\bar F_1)}^2{(b+F_1+\bar
F_2+a\,b\,F_1\bar F_2)}^2}\,.\eea
In order for the solution to be real, we shall take $a,b$ to be real
and  $F_1=F_2\equiv F$. For $a=0$, the solution reduces to the
special solution (\ref{VII.1}). When $ab^2=1$, we obtain the other
special solution (\ref{VII.2}).

It follows from 
\be \partial_{\bar 1} k_2=(K^{(2)}_0)^2\,, \ee 
that we have
\be k_2=\fft{a\,F'\bar F}{4\,(1-a\,F\bar F)}-\fft{F'(1+a\,b\,\bar
F)}{4(b+F+\bar F+a\,b\,F\bar F)} =\fft{F'(2 a^2\,b\,F\,\bar
F^2+a\,\bar F^2+2a\,\,F\,\bar F-1)}{4\,(1-a\,F\bar F)(b+F+\bar
F+a\,b\,F\bar F)}\,.\ee
Up to a gauge transformation of the K\"ahler potential, the second
equation of (\ref{VII}) implies 
\bea
\partial_{\bar 1}k_1
=2{K^{(2)}_0}\bar k_1\,.\eea
We take simplest solution of $k_1=0$, it follows that we have
\bea &&\partial_{1}\partial_{\bar 1}K^{(1)}_0=\fft{2}{\big(F'\bar
F'\big)^{\fft12}}\fft{{(1-a\,F\bar F)}{(b+F+\bar F+a\,b\,F\bar
F)}}{(a\,F^2+2\,a\,b\, F+1)^{\fft12}(a\,{\bar F}^2+2\,a\,b\, {\bar
F}+1)^{\fft12}}\,,~~~~\\&&z_2\bar z_2\,K^{(2)}_0(z_1,\bar
z_1)+K^{(3)}_0(z_1,\bar z_1,z_2)+\bar K^{(3)}_0(\bar z_1, z_1,\bar
z_2) \cr&=&{z_2\bar z_2\big(F'\bar
F'\big)^{\fft12}}\fft{{(a\,F^2+2\,a\,b\, F+1)^{\fft12}(a\,{\bar
F}^2+2\,a\,b\, {\bar F}+1)^{\fft12}}} {2{(1-a\,F\bar F)}{(b+F+\bar
F+a\,b\,F\bar F)}} \cr&&+z_2^2 F'\fft{(2 a^2\,b\,F\,\bar F^2+a\,\bar
F^2+2a\,\,F\,\bar F-1)}{4\,(1-a\,F\bar F)(b+F+\bar F+a\,b\,F\bar F)}
+{\bar z}_2^2 {\bar F}'\fft{(2 a^2\,b\,F^2\,\bar
F+a\,F^2+2a\,\,F\,\bar F-1)}{4\,(1-a\,F\bar F)(b+F+\bar
F+a\,b\,F\bar F)}\,.\nn\eea 
After the coordinate transformation $\tilde z_1(z_1)=\int
F'^{-\fft12}dz_1$ and $\tilde z_2=z_2F'^{\fft12}$ and dropping off
the tildes afterwards, the K\"ahler potential of the base space
becomes 
\bea &&K_{0}=2\,H+{z_2\bar z_2}\fft{{(a\,F^2+2\,a\,b\,
F+1)^{\fft12}(a\,{\bar F}^2+2\,a\,b\, {\bar F}+1)^{\fft12}}}
{2{(1-a\,F\bar F)}{(b+F+\bar F+a\,b\,F\bar F)}} \cr&&+z_2^2 \fft{(2
a^2\,b\,F\,\bar F^2+a\,\bar F^2+2a\,\,F\,\bar F-1)}{4\,(1-a\,F\bar
F)(b+F+\bar F+a\,b\,F\bar F)} +{\bar z}_2^2 \fft{(2
a^2\,b\,F^2\,\bar F+a\,F^2+2a\,\,F\,\bar F-1)}{4\,(1-a\,F\bar
F)(b+F+\bar F+a\,b\,F\bar F)}\,, \cr 
&&H(z_1,\bar z_1)=\int d z_1d\bar z_1\fft{{(1-a\,F\bar F)}{(b+F+\bar
F+a\,b\,F\bar F)}} {(a\,F^2+2\,a\,b\, F+1)^{\fft12}(a\,{\bar
F}^2+2\,a\,b\, {\bar F}+1)^{\fft12}}\,.\eea 
The corresponding CY2 base is given by 
\bea ds^2\!\!\!&=\!\!\!&\Big[\fft{{2(1-a\,F\bar F)}{(b+F+\bar
F+a\,b\,F\bar F)}} {H_2^{\fft12}{\bar H_2}^{\fft12}}
+\fft{|{z_2\,H_1\,H_2^{-\fft34}{\bar H_2}^{\fft14}+\bar
z_2\,H_2^{\fft34}}\,\bar H_2^{\fft34}|^2}{{2(1-a\,F\bar
F)}^3{(b+F+\bar F+a\,b\,F\bar F)}^3}F'\bar F'\Big]dz_1d\bar z_1\cr&&
+\fft{H_2^{\fft12}{\bar H_2}^{\fft12}} {2{(1-a\,F\bar F)}{(b+F+\bar
F+a\,b\,F\bar F)}}dz_2d\bar z_2
\cr&&+\fft{{z_2\,H_1\,H_2^{-\fft12}{\bar H_2}^{\fft12}+\bar
z_2\,H_2\,\bar H_2}} {{2(1-a\,F\bar F)}^2{(b+F+\bar F+a\,b\,F\bar
F)}^2}F' dz_1d\bar z_2 \cr&&+ \fft{{\bar z_2\,\bar
H_1\,H_2^{\fft12}{\bar H_2}^{-\fft12}+z_2\,H_2\,\bar H_2}}
{{2(1-a\,F\bar F)}^2{(b+F+\bar F+a\,b\,F\bar F)}^2}\bar F'dz_2d\bar
z_1 \,.\label{generalcy2}\eea 
where 
\bea H_1 &=& (a\,F^3+3\,a\,b\,F^2+3\,F)(a^2\,b\,\bar F^2+a\,\bar
F)+a\,\bar F^2+a\,b\,\bar F+a\,b^2-1\,,\cr H_2&=&a\,F^2+2\,a\,b\,
F+1\,.\eea 
The proper complex vielbein is given by 
\bea \tilde\epsilon^1&=&\fft{{\sqrt{2}(1-a\,F\bar
F)^{\fft12}}{(b+F+\bar F+a\,b\,F\bar F)^{\fft12}}}
{H_2^{\fft14}{\bar H_2}^{\fft14}} dz_1\,, \cr\tilde\epsilon^2&=&
\fft{({z_2\,H_1\,H_2^{-\fft34}{\bar H_2}^{\fft14}+\bar
z_2\,H_2^{\fft34}}\,\bar H_2^{\fft34})F'} {{\sqrt{2}(1-a\,F\bar
F)^{\fft32}}{(b+F+\bar F+a\,b\,F\bar F)^{\fft32}}}dz_1
\cr&&+\fft{H_2^{\fft14}\,\bar H_2^{\fft14}} {{\sqrt{2}(1-a\,F\bar
F)^{\fft12}}{(b+F+\bar F+a\,b\,F\bar F)^{\fft12}}}dz_2\,.\eea 
To obtain the corresponding CY3 solution, we need to solve the basic
equation 
\be
\partial^2_y G+4+\fft{2\,H_2^{\fft12}{\bar H_2}^{\fft12}}{{(1-a\,
F\bar F)}{(b+F+\bar F+a\,b\,F\bar F)}}\partial_{1}
\partial_{\bar 1}G=0\,.\ee
By construction, the equation is linear.  In fact, it implies that
$(G+2y^2)$ is the harmonic function on the space 
\be ds_3^2=dy^2+\fft{{2(1-a\,F\bar F)}{(b+F+\bar F+a\,b\,F\bar F)}}
{H_2^{\fft12}{\bar H_2}^{\fft12}}dz_1d\bar z_1\,,\ee 
which is a direct product of $\R$ and a K\"ahler 2-space. The
corresponding CY3 metric is given by 
\bea ds^2 &=&f^2
ds_3^2+f^{-2}\,\left[d\alpha+\frac{1}{4}(dx_2\partial_{x_1}-dx_1
\partial_{x_2})\partial_{x_3}
G\right]^2 +\tilde\epsilon^2\,\bar{\tilde\epsilon}^2 \,,\cr
f^2&=&1+\fft{H_2^{\fft12}{\bar H_2}^{\fft12}}{{2(1-a\,F\bar
F)}{(b+F+\bar F+a\,b\,F\bar F)}}
\partial_{1}
\partial_{\bar 1}G=-\fft14\partial^2_yG\,.\eea 
As in the previous special examples, we can rewrite the metric in
the following form
\bea ds^2 &=& V^{-1} (d\alpha + A_i dx_i)^2 + V\, ds_3^2
+\tilde\epsilon^2\,\bar{\tilde\epsilon}^2\,, \cr \triangle_3 V &=&
0\,,\qquad dV= *_3 d A\,. \eea
Written in this form, the metric is complex line bundle over ${\cal
X}_4$, where ${\cal X}_4$ is a $U(1)$ bundle over $ds_3^2$, which is
a direct product $\R$ associated with the coordinate $y$ and the
two-dimensional K\"ahler space associated with coordinates
$(z_1,\bar z_1)$.  When $ds_3^2$ becomes flat, ${\cal X}_4$
describes the Gibbons-Hawking instanton. This decomposition of the
CY3 metric is different from the original construction, where the
metric is viewed as a complex line bundle, associated with the
coordinate $(y, \alpha)$, over the CY2 base metric, given by
(\ref{generalcy2}). The CY2 metric (\ref{generalcy2}) is highly
non-trivial in that it has no Killing vector and depends on all the
coordinates; it thus lies outside of our construction for $p=1$.

\end{document}